\documentclass[12pt]{iopart}
\usepackage[dvips]{graphicx}

\begin{document}

\title{Size scaling of strength in thin film delamination}

\textbf{}

\textbf{}

\author{M. Zaiser}

\textbf{}

\textbf{}

\address{The University of Edinburgh, Center for Materials Science and Engineering, The King's Buildings, Sanderson Building, Edinburgh EH93JL, UK}

\eads{\mailto{M.Zaiser@ed.ac.uk}}

\begin{abstract}

We investigate by numerical simulation the system size dependence of the shear delamination strength of thin elastic films. The films are connected to a rigid substrate by a disordered interface containing a pre-existing crack. The size dependence of the strength of this system is found to depend crucially on the crack shape. For circular cracks, we observe a crossover between a size-independent regime at large crack radii which is controlled by propagation of the pre-existing crack, and a size-dependent regime at small radii which is dominated by nucleation of new cracks in other locations. For cracks of finite width that span the system transversally, we observe for all values of the crack length a logarithmic system size dependence of the failure stress. The results are interpreted in terms of extreme value statistics. 

\end{abstract}

\section{Introduction}

The strength of disordered materials depends on multiple characteristic length scales, ranging from the macroscopic size of the sample over the sizes of pre-existing cracks down to characteristic length scales of the disordered microstructure. Sample size dependence is commonly related to the concept of a 'weakest link': Larger samples on average contain more weak parts and thus, provided failure is governed by the weakest part of the microstructure (e.g., the largest crack), they are likely to fail at lower stress. This argument puts the size dependence of failure firmly into the realm of extreme value statistics. Samples containing a single large crack, on the other hand, fall into an entirely different category: As failure is governed by the supercritical propagation of this crack, the behavior is expected to depend on the crack size but -- at least in the large sample limit where boundary effects are negligible -- not on the size of the sample. 

In disordered quasi-brittle materials, the above arguments must be used with caution. In such materials, large flaws can emerge from the collective dynamics of interacting microcracks. Thus, the concepts of 'weakest link' or 'critical flaw' are far from straightforward -- rather than being fixed pre-existing entities, weak zones and the associated length scales may need to be envisaged as emergent features of the complex dynamics of interacting defects. At the same time, the properties of pre-existing cracks are strongly influenced by their interactions with the disordered microstructure, which may give rise to crack front roughening \cite{schmittbuhl99,zaiser09a}, the formation of self-affine fracture surfaces \cite{bouchaud03}, \cite{alava06b} and the correlated nucleation of microcracks in a process zone ahead of the crack tip \cite{alava08}.

The fracture behavior of structurally disordered materials has been studied in numerous investigations (for overview, see \cite{alava06a}). The size dependence of failure stresses in pre-cracked samples, which is in the focus of the present study, has been investigated for two-dimensional disordered sheets by Alava and co-workers \cite{alava08}. In that study, the crack was one-dimensional and contained within the plane of the sheet. In the present study, we also consider a two-dimensional system (a thin elastic film) but the crack is in the present case two-dimensional and runs parallel to the plane of the film. This can be envisaged as a delamination problem: we are dealing with a crack running along a weak interface where the film is tethered to a rigid substrate. Failure is induced by shear tractions that are applied to the free surface of the film. The considered geometry is found in several physical applications, ranging from abrasive wear of coatings and shear-induced delamination of thin films to geophysical problems such as the failure of slopes by shear band formation \cite{palmer73} and the initiation of snow slab avalanches by failure of weak interfaces in snow stratifications \cite{fyffe04,heierli08}. From a theoretical point of view, a motivation for studying this system lies in the fact that the two-dimensional crack has additional degrees of freedom which, as we shall see, may lead to behavior which differs from that of one-dimensional cracks. 

\section{Formulation of the model}

\begin{figure}[tbh]
\begin{center}
\includegraphics[width=8cm]{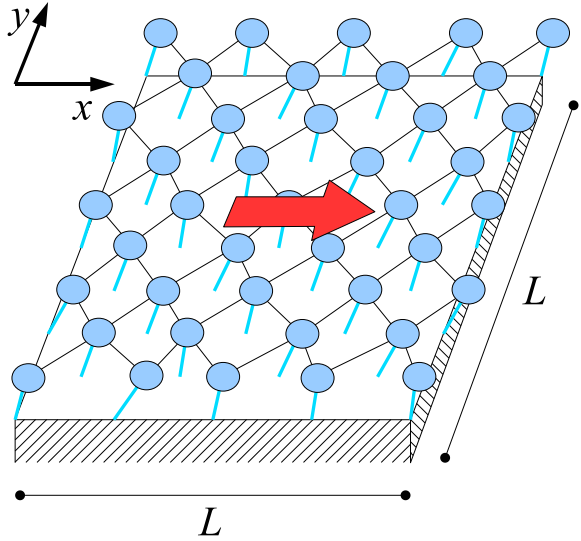}
\end{center}
\caption{Schematic illustration of the model studied in this paper: A thin elastic film is connected by a cohesive interface to a rigid substrate; the film is loaded by shear tractions applied to its free surface. The film is modelled as a regular network of identical coil springs, while the interface is represented by breakable leaf springs which connect the nodes of this network to a rigid support. }
\label{thinfilm}
\end{figure}

We consider an elastic film of thickness $D$ tethered to a rigid substrate. The film is modeled as a regular network of identical coil springs of unit length and stiffness $I$. The (discrete) coordinates of the network nodes are denoted by $X,Y$,
the network size is $L \times L$, and the origin of the coordinate system is located in the centre of the system. The interface with the substrate is represented by breakable coil springs which connect each node to a rigid support as shown schematically in Figure \ref{thinfilm}. We assume that the coil springs restrict displacements to be in the $X$ direction only, thus we are dealing with a scalar displacement field $U:=U_X$. The system is driven by an external shear force $\Sigma$ acting in the $X$ direction. This force is applied uniformly to all nodes of the network.

The leaf springs break at the uniform critical extension $U_{\rm C}=2$. Structural randomness of the interface is modelled as  quenched disorder in the form of random variations of the fracture energies $W_{\rm F}$ or, equivalently, of the leaf spring stiffnesses $K = 2 W_{\rm F}/U_{\rm C}^2 = W_{\rm F}/2$. These are assumed as independent identically distributed random variables with Weibull distribution, $P(W_{\rm F})=1- \exp[-(W_{\rm F}/W_0)^{\beta}]$. The parameter $W_0$ is chosen such that the mean value $\langle W_{\rm F} \rangle = W_0 \Gamma(1+1/\beta) = 1$ where $\Gamma(x)$ denotes the Gamma function. Different degrees of randomness can be introduced by varying the exponent $\beta$ and thus the coefficient of variation $CV = \sqrt{\Gamma(1+2/\beta)/(\Gamma(1+1/\beta))^2-1}$ of the distribution. A crack is introduced by removing all leaf springs over the cracked area. We consider two crack geometries: (i) a crack of width $2A$ in the $X$ direction that is located in the system centre and spans the system in $Y$ direction, and (ii) a circular crack of radius $A$ in the centre of the system. 

We note that a scalar model very similar to the present one was studied by Zapperi et. al. \cite{zapperi00}. The main differences between both models concern the crack geometry and driving method: Zapperi et. al,. consider straight cracks that cut from one edge of the system halfway into the interface, and drive these by imposing a growing displacement on the cracked edge of the film. This geometry could be adapted to study size effects in systems with straight cracks, but it is not well suited for circular crack geometries. 

In the limit of slowly varying displacement fields, the force equilibrium condition 
for our model can be written in the following continuum approximation:  
\begin{equation}
I \left(\frac{\partial^2 U}{\partial X^2} + \frac{\partial^2 U}{\partial Y^2}\right) + \Sigma - T(U,X,Y) = 0\;,
\label{eq:forcebalance}
\end{equation}
where $T$ is the local characteristics of the leaf springs representing the interface:
\begin{equation}
T(U,X,Y) = \left\{
\begin{array}{l}
\displaystyle
\frac{W_{\rm F}(X,Y) U}{2}, \quad U \le 2,\\[.3cm]
0,\quad U > 2.
\end{array}\right.
\label{eq:stresstrain}
\end{equation}
We note that Eq. (\ref{eq:forcebalance}) can be obtained from a long-wavelength approximation to the continuum equations describing stress equilibrium in a thin film under shear loads. The non-dimensional variables of the present 
model can thus be related to the physical and geometrical parameters of the thin film problem. The derivation has been 
given elsewhere \cite{zaiser09a,zaiser09b} and we refer the reader to those previous studies for making the link with 
'real' physical systems.

\section{Displacement fields, critical crack lengths and critical stresses for linear and circular cracks}

For analysing the behavior of cracks in finite disordered systems, it is useful to first consider the same crack geometry in 
a system without disorder ($W_{\rm F}(X,Y)=W_{\rm F}$) and of infinite extension. We envisage two crack geometries, namely (i) a linear crack of extension $2A$ in the $X$ direction which spans the system in the $Y$ direction, and (ii) a circular crack of diameter $2A$. 

We first consider a linear crack. A critical crack is characterized by the fact that the displacement on the crack line 
(in this case the lines $X=\pm A$) has the critical value $U=2$. With the additional requirement that the displacement gradient must vanish at infinity, we find that the solution of Eqs. (\ref{eq:forcebalance}) and (\ref{eq:stresstrain}) is given by
\begin{equation}
U(X)= \left\{
\begin{array}{l}
\displaystyle
\frac{\Sigma}{2 I}(A^2-X^2)+2,\quad |X|\le A,\\[.3cm]
\displaystyle
\frac{2\Sigma}{W_{\rm F}} + \left(2-\frac{2\Sigma}{W_{\rm F}}\right)\exp\left[-\sqrt{\frac{W_{\rm F}}{2I}}|X-A|\right],\quad |X| > A.
\end{array}\right.
\end{equation}
The critical crack length follows from the requirement that the displacement must be differentiable at $|X|=A$:
\begin{equation}
A = \frac{\sqrt{2W_{\rm F}I}}{\Sigma} - A^*
\label{eq:sizestrength1}
\end{equation}
where $A^* = \sqrt{2I/W_{\rm F}}$ is a characteristic length which governs the exponential decay of displacement and stress fields ahead of the crack tip. We note that the present relation
\begin{equation}
\Sigma (A+A^*) \propto \sqrt{W_{\rm F}}
\label{eq:griffith}
\end{equation}
represents a scaling that is different from Griffith's classical relation $\Sigma \sqrt{A} \propto \sqrt{W_{\rm f}}$ for bulk cracks (which also holds for linear cracks in thin sheets as studied by Alava et. al. \cite{alava08}). The different scaling reflects the fact that in case of a thin film, stress relaxation due to crack opening is confined to a volume that is limited by the film thickness (which is unity in the present non-dimensional representation). The characteristic length $A^*$ enters the expression (\ref{eq:griffith}) as an extension of the actual crack length $A$, i.e., it acts in a similar manner as the process zone size $A_p$ in Bazant's generalization of Griffith's relation to bulk cracks in semi-brittle materials \cite{bazant98}, $\Sigma \sqrt{A + A_p} \propto \sqrt{W_{\rm f}}$. 

For a circular crack we transform Eq. (\ref{eq:forcebalance}) to cylindrical coordinates. With $R = \sqrt{X^2 + Y^2}$ we have
\begin{equation}
I \left(\frac{\partial^2 U}{\partial R^2} + \frac{1}{R}\frac{\partial U}{\partial R}\right) + \Sigma - T(U) = 0\;.
\end{equation}
Again, for a critical crack we require the displacement to assume the critical value $U=2$ on the crack line $R=A$. The solution is then
\begin{equation}
U(X)= \left\{
\begin{array}{l}
\displaystyle
\frac{\Sigma}{4 I}(A^2-R^2)+2,\quad R\le A,\\[.3cm]
\displaystyle
\frac{2\Sigma}{W_{\rm F}} + \left(2-\frac{2\Sigma}{W_{\rm F}}\right)\frac{K_0(\sqrt{W_{\rm F}/(2I)}R)}{K_0(\sqrt{W_{\rm F}/(2I)}A)},\quad R > A,
\end{array}\right.
\end{equation}
where $K_n$ is the $n$th-order modified Bessel function of the second kind. The condition for matching the derivatives at $R=A$ becomes 
\begin{equation}
U'(A)= \frac{\Sigma A}{2 I} = \left(2-\frac{2\Sigma}{W_{\rm F}}\right)\sqrt{W_{\rm F}/2I} \frac{K_1(\sqrt{W_{\rm F}/(2I)}A}{K_0(\sqrt{W_{\rm F}/(2I)}A)}.
\end{equation}
This is readily analyzed in the limit $A \gg 1$ where our underlying continuum approximation is expected to be valid. The first and second order modified Bessel functions have the same asymptotic behavior and their ratio rapidly converges to unity. Thus, we obtain the following condition relating stress and critical crack radius:
\begin{equation}
A = 2\left(\frac{\sqrt{2W_{\rm F}I}}{\Sigma}-A^*\right)
\label{eq:sizestrength2}
\end{equation}
This relation is (but for a factor of 2) identical to that for a linear crack. In both cases, plots of the inverse stress vs. the critical crack length are expected to produce a straight line with positive $A$-axis intercept, $1/\Sigma = sA + a$ where $a = A^*/\sqrt{2W_{\rm F}I}=1/W_{\rm F}$. The slope of this plot is expected to be $s=1/\sqrt{2W_{\rm F}I}$ for linear and $s = 1/(2\sqrt{2W_{\rm F}I})$ for circular cracks.  

\section{Simulations of disordered systems}

In our simulations of disordered systems the coil spring constants $I$ were set to unity and leaf spring constants were  assigned randomly from a Weibull distribution with exponent $\beta$ and mean value $\langle K\rangle =1/2$ ($\langle W_{\rm F}\rangle=1$). System sizes were varied between $L=10$ and $L=160$. To introduce a linear crack of width $2A$ we simply removed all leaf springs within a strip of this width in the centre of the system. Thus the crack widths investigated are integer numbers. Circular cracks were introduced by choosing a number $B$, removing all leaf springs for which $X^2+Y^2 < B^2$, and then evaluating the 'radius' of the crack as $A = \sqrt{N/\pi}$ where $N$ is the number of removed springs. For each set of parameters $(\beta,L,A)$ we performed -- depending on system size -- typically $10^3$ simulations with different realizations of the quenched disorder. 

Simulations were performed in the standard manner for a random spring model. We solve the force equilibrium equations in the small-displacement approximation for unit applied stress and identify the leaf spring with the largest displacement $\hat{U}$. The external stress is then multiplied with $\Sigma = 2/\hat{U}$ such that this spring has the critical displacement $U=2$, and the leaf spring is removed. We then again solve the equilibrium equations and identify the leaf spring with the largest displacement. If this exceeds the critical level, this spring is also removed and the process is repeated until a new equilibrium configuration is found where $U < 2$ for all sites where leaf springs are still in place. We then again identify the leaf spring with the largest displacement, multiply the  external stress with $2/\hat{U}$, remove the spring, and so on until global failure of the interface occurs, defined in this case as the failure of {\em all} leaf springs. 

\begin{figure}[tbh]
\begin{center}
\includegraphics{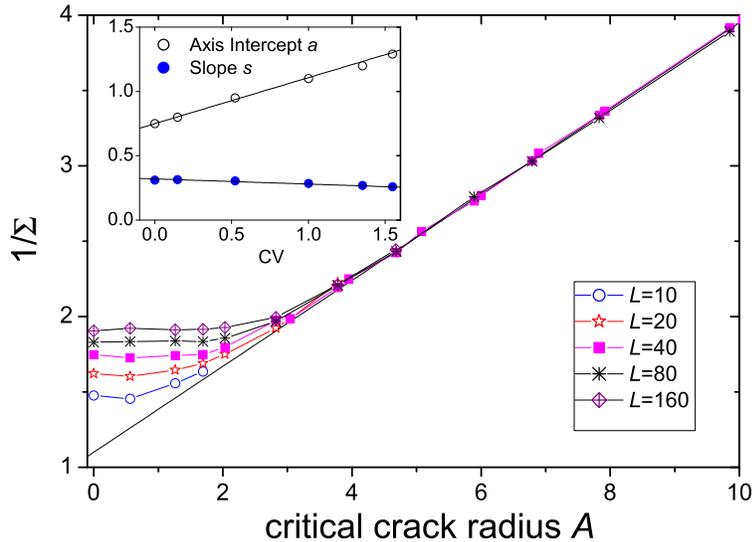}
\end{center}
\caption{System failure stress versus crack length for circular cracks in systems of different size; $\beta = CV = 1$. Inset: Dependence of slope and axis intercept of the linear regime on the coefficient of variation of the fracture energy distribution.}
\label{sizeffectcircular}
\end{figure}

\begin{figure}[tbh]
\begin{center}
\includegraphics{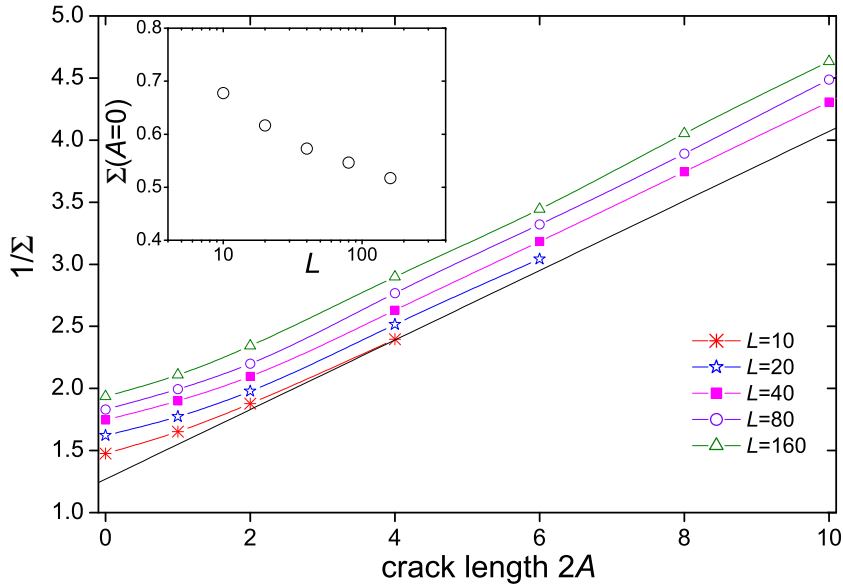}
\end{center}
\caption{System failure stress versus crack length for linear cracks in systems of different size; $\beta = CV = 1$. Inset: System size dependence of the strength of the crack-free system.}
\label{sizeffectlinear}
\end{figure}

Results are shown in Figure \ref{sizeffectcircular} for circular cracks, and in Figure \ref{sizeffectlinear} for linear cracks. For circular cracks, two different types of behavior can be observed. At small crack radii, the failure stress is approximately independent of the crack radius $A$ and has the same value as in a crack-free system. This failure stress has a logarithmic system size dependence as shown in the inset of Figure \ref{sizeffectlinear}. At large crack radii, on the other hand, the failure stress is system size independent but depends on the crack radius. This dependence is similar to what we expect for a system without disorder: the plot of $1/\Sigma$ vs. $A$ shows a linear dependence. The slope $s$ and axis intercept $a$ of this linear regime do not depend on system size. They exhibit only a weak disorder dependence as shown in the inset of Figure \ref{sizeffectcircular}. The numerical values are close to the values $a = 1$ and $s = 1/(2\sqrt{2})$ expected in the continuum limit of a disorder-free system. 

The failure stresses of systems with linear cracks shown in Figure \ref{sizeffectlinear} show different behavior. Here, the two distinct regimes manifest from Figure \ref{sizeffectcircular} are absent. Instead, the plots of $1/\Sigma$ vs. $A$ are nearly  linear for all crack sizes, and the logarithmic system size dependence of the failure stresses persists at all crack lengths. The slopes of the plots do not depend on system size. Their value is practically the same as for circular cracks, in good agreement with the deterministic result. 

\section{Discussion and Conclusions}

The behavior we observe for circular cracks is analogous to the behavior of linear cracks in thin sheets described by Alava et. al. \cite{alava08}. The cross-over between a system-size dependent but crack-size independent regime at small crack sizes, and a crack-size dependent but system-size independent regime at large crack sizes, can be understood in terms of a transition between crack-nucleation-controlled and crack-propagation-controlled failure. If the radius of the pre-existing crack is small, the stress needed for its propagation may be high enough to nucleate another critical crack in another part of a sufficiently large system. In this case, failure is not controlled by the initial crack but by crack nucleation in a 'weakest spot' elsewhere. This has been shown to give, in a thin-film geometry similar to the present one, rise to a logarithmic system size dependence of the failure stress \cite{zaiser09b}. If the radius of the pre-existing crack is large enough, on the other hand, failure is governed by propagation of this crack and does not depend on system size. The crack length dependence of the failure stress matches the expectation from deterministic fracture mechanics: We find a linear dependence of the inverse failure stress on crack length, $1/\Sigma = a + sA$. The parameters $a$ and $s$ of the linear fit exhibit a weak, approximately linear dependence on the coefficient of variation of the failure energy distribution. The slope $s$, which in deterministic fracture mechanics is proportional to $s\propto 1/\sqrt{W_{\rm F}}$, decreases slightly with increasing disorder. This indicates a disorder-induced increase of the effective fracture energy, which can be understood as a consequence of crack front roughening and crack pinning as discussed in \cite{zaiser09a}. The axis intercept $a$, by contrast, increases with increasing disorder. This behavior cannot be understood in terms of an increase in effective fracture energy. Rather, we may interpret the increase of $a$ (or of $A^*$) as a consequence of an advance of the crack beyond the initial crack line that is associated with crack front roughening under subcritical loads, an effect that becomes more pronounced as the disorder increases \cite{zaiser09a}. 

\begin{figure}[tbh]
\begin{center}
\includegraphics{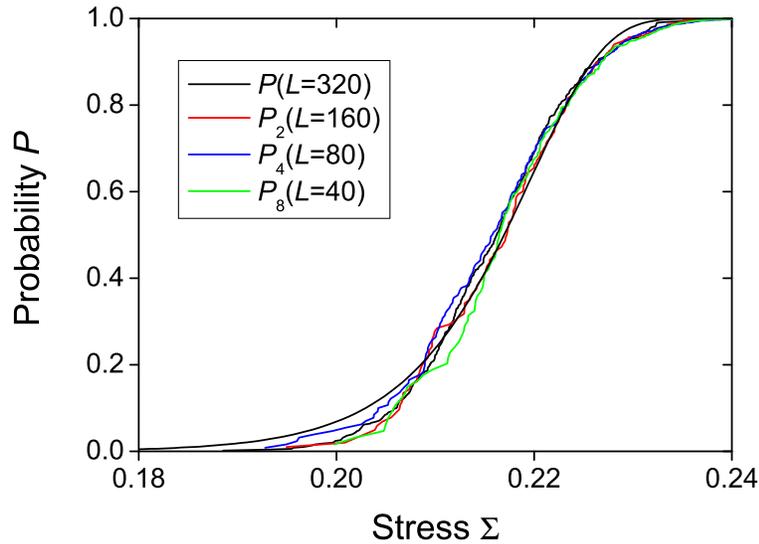}
\end{center}
\caption{Comparison of the probability distribution of failure stresses for a system of size $L=160$ with the second extremal distribution for a system of size $L=80$ and the fourth extremal distribution for a system of size $L=40$. Parameters: crack
length $2A=10$, $\beta = CV = 1$. Full line: best fit using a Gumbel distribution with parameters $C=134.42, \Sigma_0=0.2197$.}
\label{extremaldist}
\end{figure}

How can we understand the different size-dependent behavior of linear cracks? The obvious difference between the two geometries is that, for a crack spanning the entire system, the length of the crack line is equal to the system size whereas for a circular crack, it is proportional to the crack radius. As demonstrated in \cite{zaiser09a}, crack lines in interface fracture exhibit self-affine roughening, but the regime of correlated roughening is confined to scales less than the crack width $A$. Above this scale, the critical crack can envisaged as a sequence of independently pinned segments of length $\sim A$. Supercritical crack propagation can be triggered by the depinning and expansion of any of these segments. Accordingly, the system strength is governed by the weakest link along the crack line. The validity of this weakest-link argument can be directly demonstrated by analyzing the probability distributions of failure stresses: If the argument is valid, the distribution $P_L^{(n)}(\Sigma)$ of the minimum failure stress determined from $n$ independent simulations of systems of size $L$ must be equal to the distribution $P_{nL}(\Sigma)$ of the failure stress of a system of size $nL$. The $n$-th extremal distribution $P_L^{(n)}$ is calculated from the underlying distribution $P_L$ using the basic relation of extreme value statistics: $P_L^{(n)}(\Sigma) = 1 - [1- P_L(\Sigma)]^n$. Results of a comparison between $P_{nL}$ and $P_L^{(n)}$ for different values of $L$ and $n$ are shown in Figure \ref{extremaldist}. It is evident that the weakest-link argument works quite well. We note, however, that the same is not true for system sizes that are comparable to the crack length $A$ (such as $L=20$ for the parameters used in Figure 4). We attribute this to the correlated nature of the dynamics on scales $< A$, which precludes the use of simple weakest-link arguments.

Establishing the nature of the associated size effect requires understanding of the functional nature of the failure stress distribution. Fitting the observed distributions indicates that, among the stable extreme value distributions, only the Gumbel distribution $P(\Sigma) = 1 - \exp[-\exp(C(\Sigma - \Sigma_0)]$ does an acceptable job. Evidently, this implies a logarithmic size effect as observed in the simulations. 

In conclusion, the present investigation demonstrates that the additional degrees of freedom associated with 2D as compared to 1D cracks may give rise to additional types of size dependent behavior. For 2D cracks as investigated in this study, the process zone width may be associated with crack front roughening rather than with preferential accumulation of damage ahead of the crack tip. The latter mechanism, which characterizes the process zone of linear cracks in 2D sheets where it leads to screening of crack-tip stress concentrations \cite{alava08}, is irrelevant in the present case where elastic couplings are short ranged: In the present geometry, stress and displacement fields ahead of the crack decay exponentially even in the absence of disorder, hence, the crack has negligible influence on its wider surroundings. In case of linear system-spanning cracks, failure is governed by the depinning of a crack-line segment with a width of the order of the crack length. Since system failure depends on the pinning strength of the weakest segment along the crack line, this gives rise to a size dependence of the failure stresses at all crack lengths. The nature of the failure stress distributions, as well as direct determination of the mean failure stresses for different system sizes, indicate a logarithmic size effect. 

Of course, a most interesting question concerns the size-dependent behavior of 2D cracks in bulk systems where both crack front roughening and damage accumulation ahead of the crack tip are likely to operate simultaneously. The interplay of these two mechanisms, the concomitant process zone characteristics, and the resulting size effects remain a task for further studies relying on much more substantial deployment of computer power.


\Bibliography{99}
\bibitem{schmittbuhl99} Delaplace A, Schmittbuhl J, and Maloy K J,{\it High resolution description of a crack front in a heterogeneous Plexiglas block}, 1999, {\it Phys. Rev. E}, {\bf 60}, 1337-1343.
\bibitem{zaiser09a} Zaiser M, Moretti P, Konstantinidis A and Aifantis E C, {\it Roughening and pinning of interface cracks in shear delamination of thin films}, 2009, {\it J. Stat. Mech.: Theory and Experiment,} P11009.
\bibitem{bouchaud03} Bouchaud E, {\it The morphology of fracture surfaces, a tool to understand crack propagation in complex materials}, 2003, {\it Surf. Sci. Review Lett.,} {\bf 10}, 797-814.
\bibitem{alava06b} Alava M J, Nukala P K K N, and Zapperi S, {\it Morphology of two dimensional fracture surfaces}, 2006, {\it J. Stat. Mech.: Theory and Experiment,} {\bf 11}, L10002.
\bibitem{alava06a} Alava M J, Nukala P K K N, and Zapperi S, {\it Statistical models of fracture}, 2006, {\it Adv. Phys,}
{\bf 55}, 349-476.
\bibitem{alava08} Alava M J, Nukala P K K N, and Zapperi S, {\it Role of disorder in the size-scaling of material strength},  2008, {\it Phys. Rev. Letters}, {\bf 100} 0555502.
\bibitem{palmer73} Palmer A C and Rice J R, {\it The growth of slip surfaces in the progressive failure of over-consolidated clay}, 1973, {\it Proc. Royal Society A} {\bf 332}, 527-548.
\bibitem{fyffe04} Fyffe  B, and Zaiser M, {\it The effects of snow variability on slab avalanche release}, 2004 {\it Cold Reg. Sci. Techn.,} {\bf 40}, 229-242.
\bibitem{heierli08} Heierli J, Gumbsch P and Zaiser M, {\it Anticrack nucleation as triggering mechanism for snow slab avalanches}, 2008, {\it Science}, {\bf 321}, 240-243. 
\bibitem{zapperi00} Zapperi S, Hermann H J, and Roux S, {\it Planar cracks in the fuse model}, 2000 {\it Europ. Phys. J. B} {\bf 17}, 131-136. 
\bibitem{zaiser09b} Zaiser M, Moretti P, Konstantinidis A and Aifantis E C, {\it Nucleation of interfacial shear cracks in thin films on disordered substrates}, 2009, {\it J. Stat. Mech.: Theory and Experiment,} P02047.
\bibitem{bazant98} Bazant Z P and Planas J, {\it Fracture and Size Effect in Concrete
and Other Quasibrittle Materials}, 1998, CRC Press, Boca Raton (FLA). 
\end{thebibliography}
\end{document}